\documentclass[5p,10pt,sort&compress]{elsarticle}
\usepackage{multirow}
\usepackage{ragged2e}
\usepackage{ulem}
\usepackage{todonotes}
\usepackage{xcolor}
\usepackage{hyperref}
           
\journal{
 arXiv
 }

\begin{document}

\begin{frontmatter}
\title{\textbf{Effect of lattice shrinking on the migration of water within zeolite LTA}}
\author[addUPO]{Julio E. Perez-Carbajo}
\author[addUPO]{Salvador R. G. Balestra}
\author[addUPO,addTUe]{Sofia Calero\corref{cor1}}
\ead{scalero@upo.es}
\author[addUPO]{Patrick J. Merkling\corref{cor1}}
\ead{pjmerx@upo.es}
\address[addUPO]{Department of Physical, Chemical, and Natural Systems, Universidad Pablo de Olavide, Ctra. Utrera km 1, 41013 Seville, Spain}
\address[addTUe]{Technische Universiteit Eindhoven, 5600 MB Eindhoven, The Netherlands}
\cortext[cor1]{Corresponding author}


\begin{abstract}
Water adsorption within zeolites of the Linde Type A (LTA) structure plays an important role in processes of water removal from solvents. For this purpose, knowing in which adsorption sites water is preferably found is of interest. In this paper, the distribution of water within LTA is investigated in several aluminum-substituted frameworks ranging from a Si:Al ratio of 1 (maximum substitution, framework is hydrophilic) to a Si:Al ratio of 191 (almost pure siliceous framework, it is hydrophobic). The counterion is sodium. In the hydrophobic framework, water enters the large $\alpha$-cages, whereas in the most hydrophilic frameworks, water enters preferably the small $\beta$-cages. For frameworks with moderate aluminum substitution, $\beta$-cages are populated first, but at intermediate pressures water favors $\alpha$-cages instead. Framework composition and pressure therefore drive water molecules selectively towards $\alpha$- or $\beta$-cages.\\

\begin{center}
\includegraphics[width=0.45\textwidth]{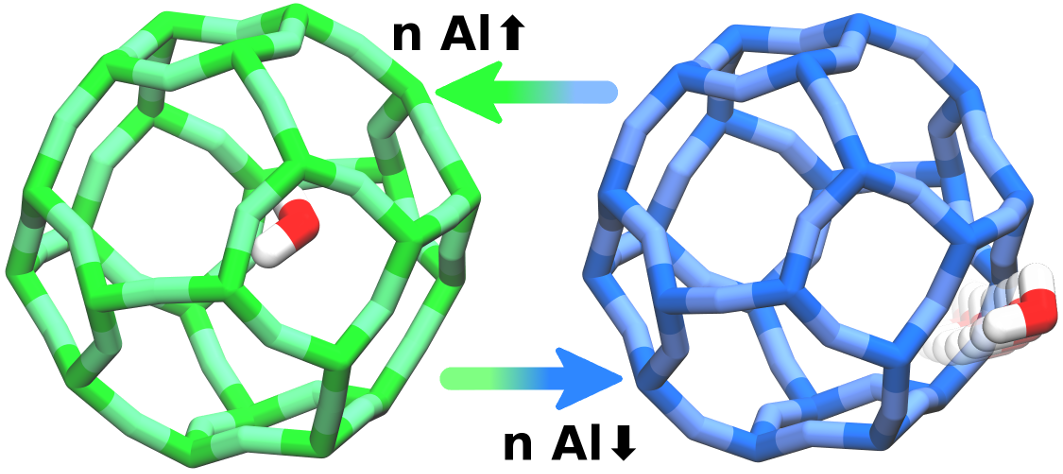}
\end{center}
\textbf{Highlight:} LTA composition and pressure allow us to drive water molecules into lta or sod-cages.
\end{abstract}

\begin{keyword}
cations, Linde Type A, water, hydrophilicity, zeolite A, NaA, adsorption
\end{keyword}

\end{frontmatter}

\section{Introduction}
Zeolites are natural or synthetic crystalline compounds containing most
commonly only silicon, aluminum, oxygen, and exchangeable cations. Zeolites
have important industrial applications due to their nanoporosity. The zeolite
of interest in this work, Linde Type A (LTA) zeolite has a cubic unit cell. It
possesses two types of roughly spherical cavities, lta cages (or
$\alpha$-cages) with an approximate diameter of 11.2 \AA~ and sodalite (sod)
cages (or $\beta$-cages) with an average diameter of 6.6 \AA. $\alpha$-cages
are connected to another six $\alpha$-cages through eight-membered windows
(S8R) of about 4.2 \AA~ and connected to eight $\beta$-cages through
six-membered windows of about 2.2 \AA~ openings. Although the idealized cell
found in the pure silica version contains one $\alpha$- and one $\beta$-cage,
and has a chemical formula of Si$_{24}$O$_{48}$, for the sake of simplicity we
will refer to the unit cell as the supercell found in aluminum-substituted
versions of chemical formula  Na$_x$[Al$_x$Si$_{192-x}$O$_{384}$] which
contains eight $\alpha$- and eight $\beta$-cages. Its side length, depending
on aluminum content, varies between 23.75--24.55 \AA~ at room temperature.

Water molecules are able to enter zeolite LTA, which can be made more
hydrophilic by substituting some of the silicon atoms by aluminum. This has
significant industrial consequences, as pervaporation processes for removal of
water using LTA have been conducted both in the lab and in large-scale
industrial plants \citep{Wenten2017}. Thus, Mitsui Engineering and
Shipbuilding Co., Japan, built an industrial facility for ethanol dehydration
using sodium-containing LTA zeolite membranes. Other applications include
removal of water from other solvents, water desalination and water removal
from esterification processes. It is therefore useful to get a more
fundamental understanding of water in LTA. How water behaves within the LTA
crystalline structure has been studied in a number of ways: by X-ray diffraction \citep{MeierXRD}, Fourier-transform infrared spectroscopy \citep{CrupiFTIR1,CrupiFTIR4}, spin-echo nuclear magnetic resonance \citep{PaoliQENSNMR}, and several neutron scattering techniques \citep{CorsaroQENS1,PaoliQENSNMR,CorsaroENS2,CorsaroIINS3}. Theoretical studies, with their very detailed insight into the microscopic world of the structure, contribute also significantly to the understanding of these systems. Adsorption isotherms \citep{Calero2014,Castillo2013,Gomez-Alvarez2015}, hydrogen-bond statistics \citep{Calero2014,Coudert2006,Coudert2009,Gomez-Alvarez2015,Gomez-Alvarez2016}, diffusion coefficients \citep{Demontis2010,Faux1999}, thermodynamic considerations \citep{Higgins2002} and some characterization of the bonding sites and behavior of water \citep{Coudert2006,Coudert2009,Demontis2008,Faux1997} have been obtained for specific compositions like pure silica LTA or the maximally aluminum-substituted LTA called LTA 4A, NaA or zeolite 4A. However, the composition-dependent location of water molecules has not been explored. Thus, it is the purpose of this work to study the distribution of water molecules in LTA across the whole range of sodium-compensated aluminum substitutions, namely from almost pure silica LTA to Na$_{96}$[Al$_{96}$Si$_{96}$O$_{384}$].

\section{Computational methods}

Structures are defined as rigid frameworks with static partial charges \citep{Garcia-Sanchez2009}, where oxygen and silicon atoms have charges of $q_{\rm{O}}=-0.3930$ $e^-$ and $q_{\rm{Si}}=+0.7860$ $e^-$, respectively.
Since introducing aluminum affects the partial charge on silicon, its charge is set to $q_{\rm{Al}}=+0.4860$ $e^-$ \citep{Jaramillo1999,Calero2004}. This charge redistribution results in Na$^+$ cations tending to be located nearby Al atoms,
affecting thereby the partial charge of these oxygen atoms forming the AlO$_4$ tetrahedra \citep{Jaramillo1999}: ($q_{\rm{O_a}}=-0.4138$ $e^-$). 
Na$^+$ extra-framework cations introduced in the structure are considered as point charges $q_{\rm{Na^+}} = +0.3834$ $e^-$ \citep{Garcia-Sanchez2009} and are allowed to move trough the system. The water molecule is defined by the TIP5P/Ew model \citep{Rick2004}. This model, formed by five sites arranged tetrahedrally, in which the oxygen atom transfers its negative charge to two \textit{dummy} pseudo-atoms, has been previously reported to reproduce the adsorption behavior of water in zeolites \citep{Castillo2009,Calero2014,Gomez-Alvarez2015,Gomez-Alvarez2016}.

Interactions between the interaction sites of the system (lattice atoms, extra-framework cations, and adsorbates) are ruled by Coulombic potential for electrostatic interactions, using the Ewald summation to handle the periodicity of the system, and Lennard--Jones (L--J) potentials are used to model van der Waals (vdW) interactions.
L--J interactions of adsorbates with zeolites are dominated by dispersive forces with the oxygen atoms (O and O$_a$) so the interactions with silicon and aluminum atoms are neglected \citep{Bezus1978,Kiselev1985}. vdW interactions are also not considered between sodium cations themselves due to their strong electrostatic interactions. The rest of L--J interactions are already parametrized in and taken from previous works \citep{Castillo2009,Calero2014}.

Different LTA-type lattices have been considered in this work, attending to their aluminum content. All of the frameworks are charge-compensated by introducing an equal number of Na$^+$ cations as aluminum atoms in the framework. LTA zeolites have been synthesized over a wide range of Si:Al ratios, from pure silica framework \citep{Fyfe1984,Corma2004} up to LTA 4A \citep{Reed1956,Pluth1980} with the same amount of Si atoms as Al atoms. This latter structure meets the theoretical maximum substitution of Si by Al atoms allowed, according to L\"{o}wenstein's rule \citep{Loewenstein1954}. The Si:Al ratios of our structures span this range, from Si:Al=1 (96 Al/Na$^+$ pairs per supercell, henceforth called ``LTA-96'' for simplicity) down to almost pure silica LTA (Si:Al=191, 1 Al/Na$^+$ pair per supercell, ``LTA-1''). The other ratios used in this study are 1.02 (95 Al/Na$^+$ per supercell, ``LTA-95''), 1.91 (66 Al/Na$^+$ per supercell, ``LTA-66''), 3.57 (42 Al/Na$^+$ per supercell, ``LTA-42''), and 5 (32 Al/Na$^+$ per supercell, ``LTA-32'').

While atomistic positions for both aluminum atoms and sodium cations are described for LTA 4A \citep{Pluth1980} and were taken from the literature, the rest of LTA structures were generated computationally. To that end, starting from the lattice of LTA 4A, aluminum atoms were progressively substituted by silicon atoms. The first substitution was made ramdomly and subsequent substitutions were restrained by Dempsey's rule \cite{Dempsey1968}, to minimize the number of Al--O--Si--O--Al elements and in order to obtain a more uniform aluminum distribution in the lattice. This method generates frameworks with well-defined properties. 

As was mentioned previously, frameworks were considered rigid throughout the simulation with the exception of the extra-framework cations but, for each Si:Al ratio, we considered two lattices. For the first one, atomistic positions of the LTA 4A lattice \citep{Pluth1980} were kept unchanged. For the second one, not only lattices, but also extra-framework cations, were allowed to relax their crystallographic positions to meet a minimum energy configuration. These minimizations were performed ten times independently and the lowest energy configuration was selected to avoid false minima and energy saddle points. This configuration was taken as the initial configuration of the simulations.

To compute adsorption isotherms of water and its average occupation profile in LTA-type zeolites, Monte Carlo simulations are run in the Grand Canonical ensemble ($\mu VT$) \citep{Frenkel2002}. Setting the chemical potential $\mu$, the fugacity of a gas $f$ and therefore the pressure are also set. Fugacity and chemical potential are related through the equation $\mu=\mu^0+RT\ln(f/p^0)$, in which $p^0$ is the standard chemical pressure, $R$ the gas constant and $T$ the temperature, set to 298K in this study. Cations are placed inside the structures using random trial insertions to bypass energy barriers \citep{Calero2004} and move by trial displacements. Since the L--J potential cutoff was set to 12~\AA~ in the development of the interaction potentials, the same cutoff has been applied in our study. The sides of our simulation boxes were at least twice the L--J cutoff.

Structural relaxations have been carried out using the GULP code \citep{Gale1997a}. We have used the well-known shell-model potentials of Sanders \textit{et al.} \citep{Sanders1984} for the structure and the potential of Jackson \textit{et al.} for the cations \citep{catlow_potential_1987,catlow_potential_1988}. Broyden--Fletcher--Goldfarb--Shanno (BFGS) and Rational Function Optimization (RFO) minimization methods were used to ensure convergence to true energy \citep{fletcher1987practical,Byrd2000}. Although the BFGS algorithm is faster than RFO, RFO behaves better than BFGS in the vicinity of the minimum. So, we have used initially the BFGS algorithm, and when the gradient norm dropped below 0.03, we have switched to the RFO minimizer. This methodology has been validated in many previous works and provides cell parameters and realistic crystal structures \citep{AlmoraBarrios2001,Lewis2002,Balestra2015}. MC simulations of this work were performed using RASPA software \citep{Dubbeldam2015}. Average occupation profiles were obtained by using the software SITES-ANALYZER \citep{AOP2017}.

\section{Results and discussion}

\begin{figure}
    \includegraphics[width=0.45\textwidth]{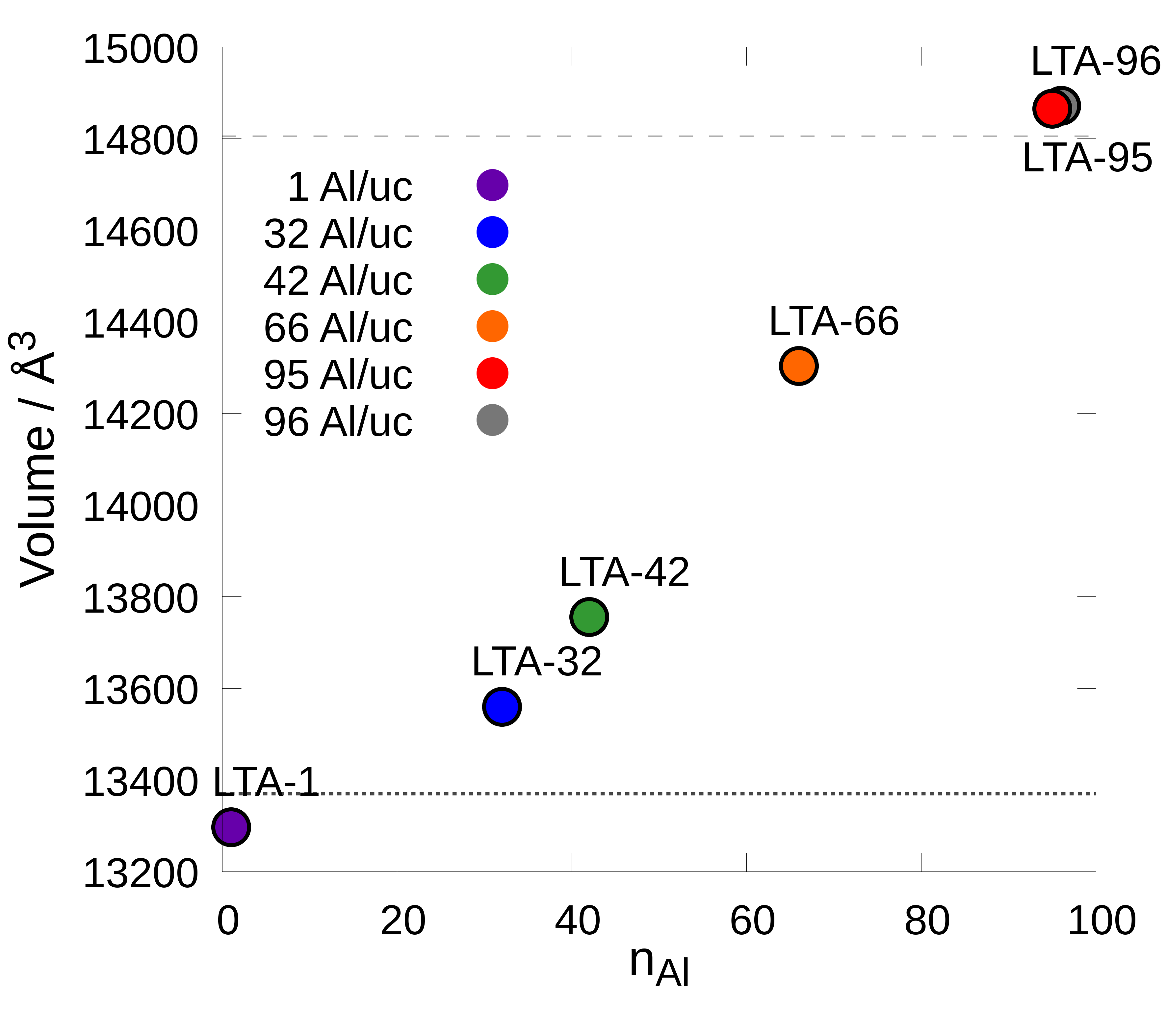}
  \caption{Minimized LTA cell volumes. Dashed line for experimental volume of LTA4A and of fixed-size frameworks. Dotted line for volume of pure silica LTA (ITQ-29). All volumes apply to $2\times2\times2$ supercell.}
    \label{img:f1}
\end{figure}

Experimentally, from purely siliceous LTA to LTA 4A the crystal cell dimensions increase noticeably \citep{Corma2004,Pluth1980}, which translates into a 10\% volume increase. This has important consequences for the pore volume of the nanoporous cavity. The force field used in our zeolite is able to reproduce the experimental volumes, as shown in Figure \ref{img:f1}. We therefore expect it to predict reasonably well the volumes of structures with an intermediate aluminum content. All of the structures were optimized as dehydrated frameworks and then frozen. This is a valid approach because in test calculations we have found the effect of hydration on volume change to be negligible. A similar observation had been done previously for LTA 4A \citep{Castillo2013}, using a different force field from ours. The size of the unit cell, and especially the size of the openings in the sodalite cages is an important factor for the capacity of the zeolite for separating multicomponent mixtures. For the sake of naming the frameworks simply, they will be called LTA-$x$, with $x$ the number of aluminum atoms in the $2\times2\times2$ supercell, i.e. the cell that contains 8 $\alpha$-cages and 8 $\beta$-cages. Thus, LTA-96 is LTA 4A.

\begin{figure}
    \includegraphics[width=0.45\textwidth]{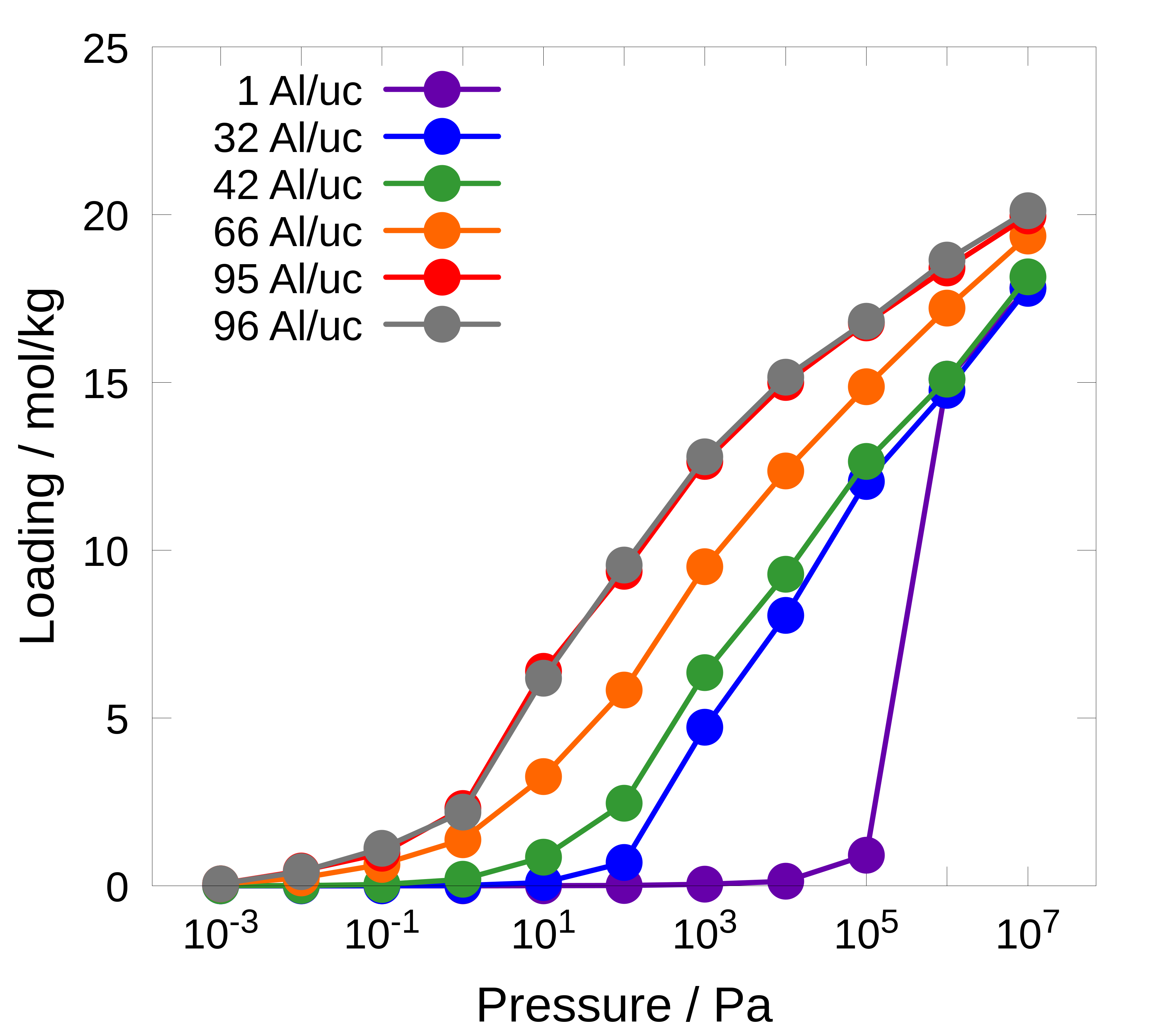}
    \caption{Full circle, solid line: Reversible water adsorption
      isotherms in minimized LTA frameworks}
    \label{img:f2}
\end{figure}

The hydrophobicity of the zeolite has dramatic consequences on adsorption behavior. The almost pure siliceous zeolite LTA-1 requires pressures in excess of 10$^5$ Pa to adsorb water (Figure \ref{img:f2}), and the adsorption curve is very steep. This is a consequence of the hydrophobic environment, but once water enters, it creates nucleation sites for other water molecules around it. Clusters of water are created. At LTA-32 already, nucleation sites exist (the sodium cations), which draw in water more gradually. Therefore, half-loading is achieved at around 10$^4$ Pa. This means that the framework at this level of aluminum-enrichment has already a marked hydrophilic character. Further aluminum enrichment of the framework to 66 aluminum atoms per supercell reduces the necessary pressure another order of magnitude, and for LTA 4A, half-loading is achieved at 10$^2$ Pa. In terms of the IUPAC classification of isotherms \citep{IUPAC_Isotherms2015}, siliceous zeolite is a type V isotherm whereas the other isotherms are of type I.

\begin{figure}
    \includegraphics[width=0.45\textwidth]{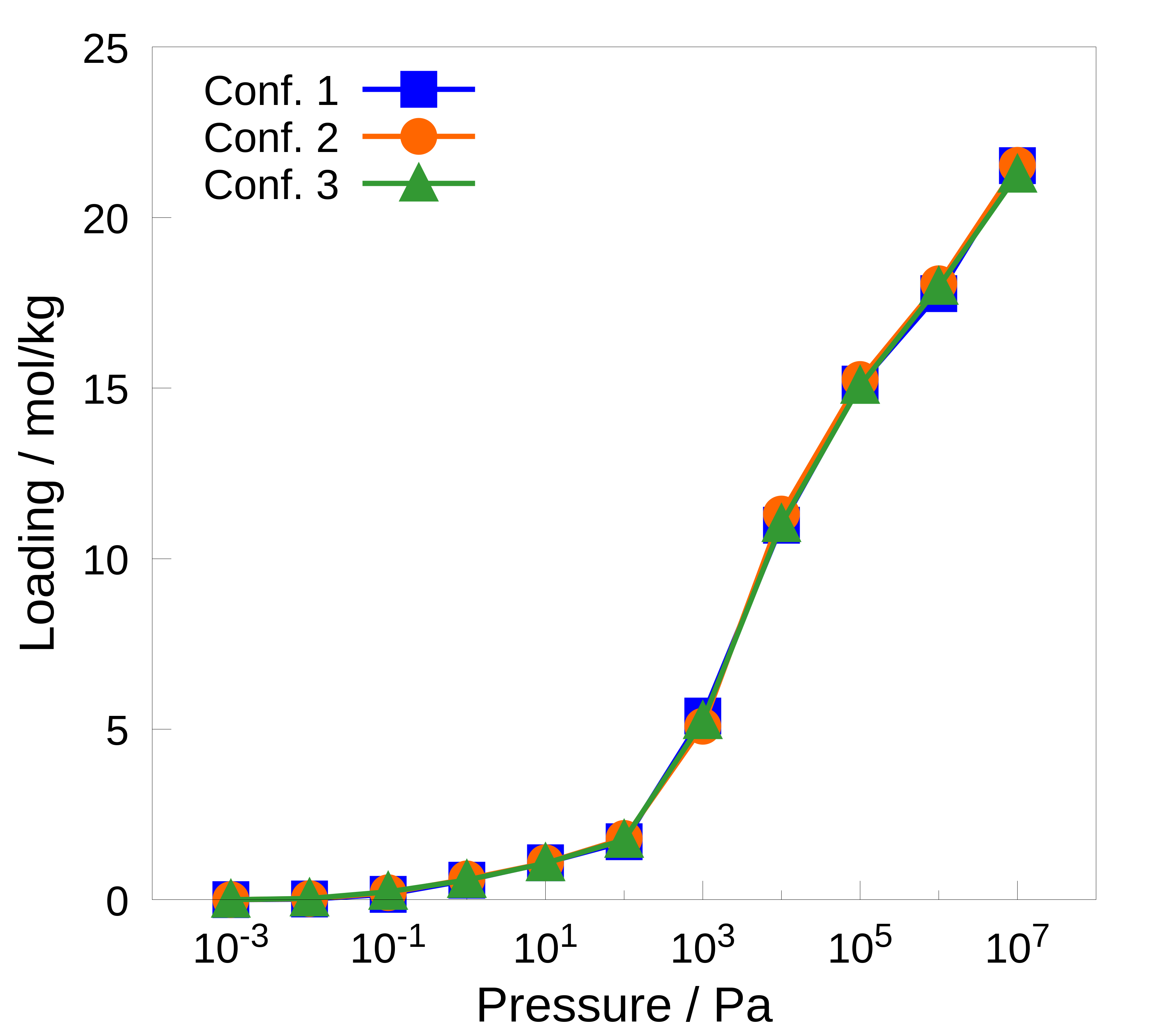}
  \caption{Water adsorption isotherms for three frameworks with the same overall composition Na$_{42}$[Al$_{42}$Si$_{150}$O$_{384}$]}
    \label{img:f3}
\end{figure}

Figure \ref{img:f3} shows the water adsorption isotherms at 298~K of three frameworks of composition LTA-42, but with a different distribution of aluminum atoms. All three distributions comply with L\"owenstein's and Dempsey's rules. It can be appreciated that the three frameworks led to virtually the same adsorption curves.

\begin{figure}
    \includegraphics[width=0.45\textwidth]{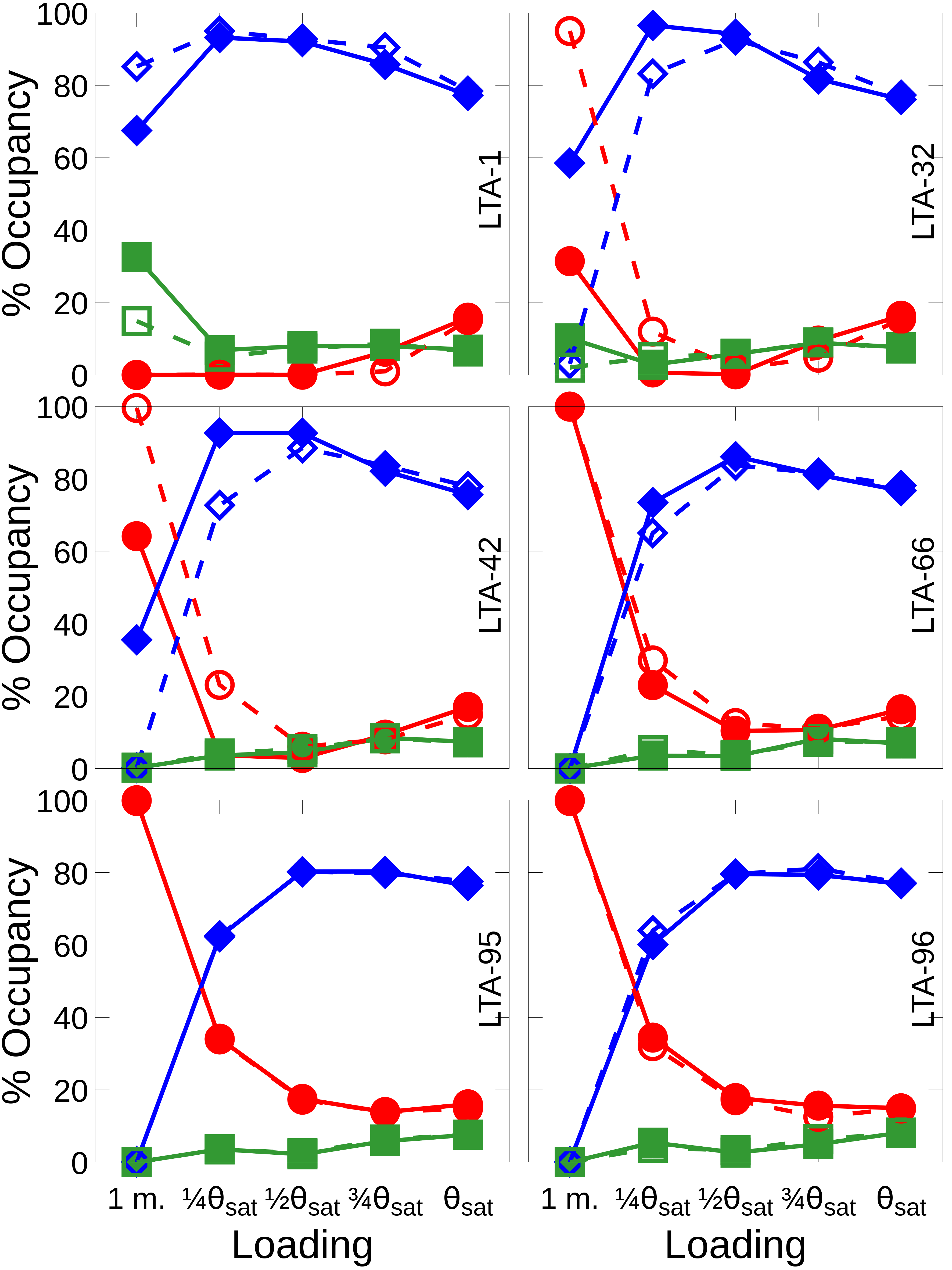}
  \caption{Distribution of water molecules by adsorption sites in percent: $\alpha$-cage (blue diamonds), $\beta$-cage (red circles) and S8R windows (green squares) for fixed-size (empty symbols, dashed lines) and minimized (full symbols, solid lines) LTA as a function of the fraction of maximum loading}
    \label{img:f4}
\end{figure}

An understanding of the site-dependent hydration can be drawn from a loading-dependent representation of the percentage of water present in the structural features of zeolite LTA, namely the small $\beta$-cages, the large $\alpha$-cages and the window-area S8R. This is shown in Figure \ref{img:f4}. At low aluminum-content, water is not contained in the $\beta$-cages until high loading, a conclusion in qualitative agreement with a study by Coudert \textit{et al.} \citep{Coudert2006} At maximum loading however, we can assume water molecules to be distributed among the sites with no preference for any given site, which means that the distribution should reflect the volumes of the regions: $\alpha$-cages account for roughly 78\% of the available pore volume, $\beta$-cages for 15\% and the S8R window area for 7\%. According to this, at lower pressures and loading, water has a strong affinity for the window region. At increasing pressure, it populates much more strongly the $\alpha$-cages, whereas $\beta$-cages are the last to be populated. Already in LTA-32, the behavior is radically different: $\beta$-cages are disproportionately populated at low pressures, even more so than the window region. At high pressures, saturation is responsible for the filling of the pores and the volume-based distribution indicated earlier is reached. $\beta$-cages contain three to four water molecules each, $\alpha$-cages 22 and three quarters of the windows one molecule. But the most striking part is the behavior at intermediate pressures, in this case 10$^3$-10$^5$ Pa, in which $\beta$-cages are statistically depopulated, on average one water molecule in the supercell, compared to of the order of 100 in the $\alpha$-cages. LTA-42 and LTA-66 exhibit a similar phenomenon. If we relate this to the sodium positions, these striking results make sense: in LTA-1, the sodium cations are located in the S8R region at low pressures, but are forced into the $\alpha$-cages at higher pressures. In LTA-32 at low pressure, sodium cations are located mainly in the window-region, and also disproportionately in the $\beta$-cages (16 and 7 Na$^+$ per window- and $\beta$-cage region of the supercell, respectively). Then, at 10$^3$ Pa, these numbers drop to 1 and 0.6 respectively, most cations and water molecules are then located in the $\alpha$-cages. From 10$^6$ Pa on, a few cations (and water molecules) are back in the $\beta$-cages. This is a very interesting behavior, because it means that by choosing the Si:Al ratio and regulating the pressure, one can direct the water towards one or another type of site. Similar patterns are seen in LTA-42 and to a lesser extent in LTA-66, sodium populations in the $\beta$-cages decrease at the intermediate pressures at which water also is driven out of these cages and gets back in again at higher pressures. No such behavior is observed in LTA-95 and LTA-96, sodium cation populations are roughly constant throughout the pressure range and water molecules are located overwhelmingly ($>$~99\%) in the $\beta$-cages at low pressure. At higher pressures, water molecules also populate $\alpha$-cages, a finding already pointed out by Castillo \textit{et al.} \citep{Castillo2013}. The curves for LTA-96 (maximum aluminum substitution) and LTA-95 (one aluminum short of maximum aluminum-substitution) are, not surprisingly, very similar, an indication for good sampling.

In the frameworks with intermediate aluminum substitution such as LTA-32, LTA-42 and LTA-66, at low pressures sodium-water interactions and interactions with the framework in the confined area of the $\beta$-cages have energetic advantages. Then, at higher loading, confinement turns into an obstacle, large clusters of water that can be formed only in the $\alpha$-cages engage in many water-water hydrogen bonds, which is energetically favorable. In the frameworks LTA-95 and LTA-96, the high amount of sodium cations relative to water means that most water molecules are part of the hydration shell of sodium, which is a stronger interaction than that provided by hydrogen bonds. 

So far, we have commented on the curves of the minimized structures in Figure \ref{img:f4}. These structures describe the confined environment and its size accurately. We have identified a behavior of sodium and water distribution that is dependent on the Si:Al ratio. But, given that the preference of a water molecule for the confined region of $\beta$-cages or the larger $\alpha$-cages is highly sensitive to the cell size, we would like to identify if the adsorption behavior at specific sites is due to the loss of charges and hydrophilicity brought about by decreasing the number of sodium and aluminum atoms, or if the shrinking of the cell with decreasing number of sodium and aluminum atoms is essential to the location of the adsorbates. To answer this question, GCMC simulations in a cell of the size of LTA 4A have been performed. The distribution of water molecules by adsorption sites is also represented in Figure \ref{img:f4}. The curves for LTA-96 and LTA-95 are almost superimposable with the ones for the minimized structures because the cell size is identical or virtually identical. Differences are due to statistics because they were obtained in independent simulation runs. The less aluminum in the structure, the greater the difference in volume (Figure \ref{img:f1}) between the fixed-size cell and the minimized cell. Some differences in water distribution show at low loading. The interpretation is that the shrinking of the cell brought about by lowering the aluminum content does not favor water population of the $\beta$-cages. The fact that water in the minimized structures LTA-32 and LTA-42 still populate disproportionately the $\beta$-cages is because the presence of cations in these cages is a powerful driver that offsets the shrinking. Qualitatively, the migration out of the $\beta$-cages and back in as pressure is increased (infinitely slowly) is also observed in the fixed-size cells. It is thus an effect of electrostatics, i.e. hydrophilicity/hydrophobicity, and not an effect of the size of the structure due to the Si:Al ratio.

\section{Summary and Conclusions}
\begin{figure}
    \includegraphics[width=0.47\textwidth]{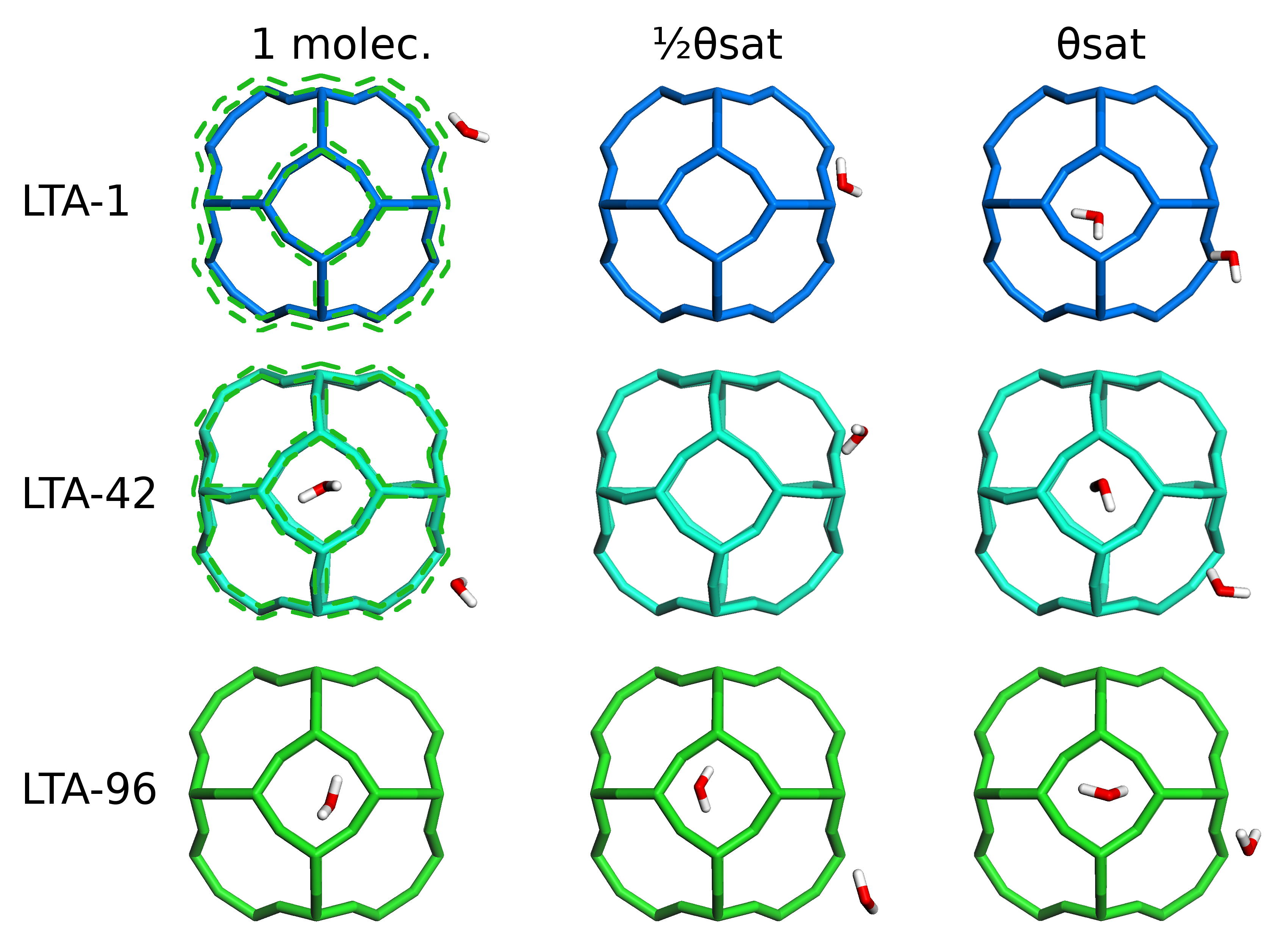}
  \caption{Schematic representation to illustrate the water migration in or out of $\beta$-cages observed in zeolite LTA depending on composition and loading. Green dashed lines in first column mark the contour of LTA-96, evidencing lattice shrinking when Si:Al ratio grows.}
    \label{img:f5}
\end{figure}

In nearly pure siliceous zeolite LTA, water is located in the $\alpha$-cages and, as pressure or loading increase, it finally gets into the $\beta$-cages. On the contrary, for frameworks with significant aluminum content (Si:Al $\leq$ 5) such as LTA-42, $\beta$-cages are populated disproportionately at low pressure. For these systems, the percentage of water in $\beta$-cages drops as more and more water gets into the $\alpha$-cages with increasing pressure. The most exciting situation arises in the systems with Si:Al ratios of 1.91-5: water molecules and sodium cations move out of the $\beta$-cages at intermediate pressures, and are forced back in at sufficiently high pressures (10$^6$ Pa), at which the whole of the available volume is occupied. These findings are summarized visually in Figure \ref{img:f5}.

This is a very interesting behavior, because it means that by choosing the Si:Al ratio and regulating the pressure, one can direct the water towards one or another type of site. This could be used technologically, because it would allow separating multi-component mixtures by tuning the adsorption selectivity of water.

\section*{Acknowledgements}

  This work was supported by the Spanish Ministerio de Econom\'{\i}a y Competitividad (MINECO) through the grant number CTQ2016-80206-P. We thank C3UPO for the HPC support. S.R.G.B. thanks Spanish Ministerio de Econom\'{\i}a y Competitividad (MINECO) for his predoctoral and postdoctoral fellowship (BES-2014-067825 from CTQ2013-48396-P).

\bibliographystyle{elsarticle-num}
\bibliography{Article_Perez-Carbajo_water_LTA}

\end{document}